# Possible canonical distributions for finite systems with nonadditive energy

Congjie Ou[1,2,5], Wei. Li[1,3], Jiulin Du[4], François Tsobnang[1],

Jincan Chen[5], Alain Le Méhauté[1], and Qiuping A. Wang[1]

[1]Institut Suprieur des Materiaux du Mans,44, Avenue F.A. Bartholdi, 72000 Le Mans, France

[2]College of Information Science and Engineering, Huaqiao University, Quanzhou 362021, People's Republic of China.

[3]Institute of Particle Physics & Complexity Science Center, Hua-Zhong Normal University, Wuhan 430079, P.R. China

[4]Department of Physics, Tianjin University, Tianjin 300072, China

[5]Department of Physics, Xiamen University, Xiamen 361005, China

Abstract

It is shown that a small system in thermodynamic equilibrium with a finite thermostat can have a q-exponential probability distribution which closely depends on the energy nonextensivity and the particle number of the thermostat. The distribution function will reduce to the exponential one at the thermodynamic limit. However, the nonextensivity of the system should not be neglected.

## 1. Introduction

The studies in finite systems such as nanometric particles, nuclei, atomic clusters as well as gravitational systems (see for example [1] and [10] and references there-in) have brought renewed interest in small systems whose statistical physics and thermodynamics are more complicated than those of large systems. A system can be called small when its size is comparable with the interaction scale between its elements. One character of the complexity of a small system is that it may be *nonextensive*, i.e., its macroscopic quantities may not be proportional to its size. They can also become *nonadditive*, meaning that if you divide a system into smaller subsystems, a thermodynamic quantity of the total system is not necessarily the sum of the same quantities of the subsystems. A possible reason for this is the non negligible surface effect of the three-dimensional (3D) system. In other words, when a small 3D system splits into smaller subparts, not only the volume change but also the surface change and surface interaction must be considered, which may yield the non proportionality of the quantity such as energy and entropy to volume or to element number of the system. Another reason may be the interactions between the subsystems. This interaction can be dismissed when the thermodynamics of the subsystems is considered separately but must be taken into account for the total system.

Another complexity of the small system arises from the fact that it has many more fluctuations than the large one. This may cause considerable difficulty in its treatment. As examples of this complexity, we can cite a possible violation of the second law of thermodynamics by nonequilibrium small system (micrometer size) within short time period (one or two seconds) [11,12] and the plausible negative specific heat of nuclear fragments[13].

Recently, many theoretical results have been published on the statistical properties of small systems and have raised questions and controversies [1,2,3,4,10]. A main point is whether or not a small system follows the Boltzmann-Gibbs statistical mechanics [1]. Among the published results, mathematical proofs from first principles have been given[2,3] to confirm that a small system, in thermodynamic equilibrium with a finite heat bath, may *a priori* have the *q*-exponential distribution of the nonextensive statistical mechanics (NSM) deduced from Tsallis entropy [5]. In our opinion, the physical consideration and the mathematical proofs are convincing to show the connection between the finiteness of the considered system and the nonextensivity of the theory. As a matter of fact, these proofs are not new. They can be found in standard textbooks on statistical mechanics (see for example [6]).

A common character of these proofs is that additive energy is used everywhere as a first hypothesis. Perhaps this approximation has been proposed in order to simplify the calculations aiming to get the statistical theory for large system [6]. This can be acceptable since with thermodynamic limits additive and extensive energy is assumed everywhere. However, if we want to address a small system, this hypothesis is questionable. The additive energy for NSM and the relative problems and controversies raised in the establishment of thermodynamic laws have been extensively discussed recently. We make no comment here. The reader can find different points of view on this topic in [7] and references there-in.

In the present work, we show that the same recipe to prove the connection of NSM distribution function to a small system is also valid with certain models characterizing the nonadditivity of energy. And it is unnecessary to use additive energy for the small system to have the distribution of NSM. The rest of this paper is organized as follow: In section 2 we introduce a simple form of the energy composition of two subsystems with the help of a classical self-gravitation system. Based on this assumption the probability distribution function is derived in section 3 and the properties of the distribution are discussed in section 4 for some different cases. In section 5 we make a summarization and some novel results are given.

## 2. Energy composition for two subsystems

We consider an adiabatically and mechanically isolated system $\Sigma$ containing finite $N$ classical particles, with the assumptions of equiprobable microstates and of ergodicity for this system [6], the distribution is given by the microcanonical one

$$p(X) = \frac{1}{\Omega(E)} \delta[E - H(X)], \qquad (1)$$

where $H(x)$ is the Hamiltonian of $\Sigma$ having energy $E$, $\Omega$ the total volume of the phase space points satisfying $H(X)=E$ and $X$ the phase space coordinates of the $N$ particles ($dX = \prod_i d\vec{r}_i d\vec{p}_i$ with $i=1,2...N$ with $\vec{p}_i$ the momentum and $\vec{r}_i$ the coordinates of the $i$th particle). Now let us divide this system into two interacting subsystems $\Sigma_1$ and $\Sigma_2$ with respectively $N_1$ and $N_2$ particles and hamiltonians $H_1(X_1)$ and $H_2(X_2)$. We suppose

$$H(X)=H(X_1, X_2)=H_1(X_1)+H_2(X_2)+U_{12}(X_1, X_2) \tag{2}$$

where $U_{12}$ is the interaction energy between $\Sigma_1$ and, say, its thermostat system $\Sigma_2$.

In the previous works[2,3,6], the derivation of the canonical statistics from the microcanonical ensemble of $\Sigma$ has been done with negligible $U_{12}$. However, as mentioned above, for small systems, $U_{12}$ may be very important even if the interaction is of short range [8]. In what follows, we will show that the same result of [6], i.e., the existence of $q$-exponential distribution for small systems, is also valid with a nonadditive energy. For this purpose, we suppose in this work that $U_{12}$ can be modeled by a simple composition of $H_1(X_1)$ and $H_2(X_2)$, i.e.,

$$U_{12}(X_1, X_2)=\lambda H_1(X_1)H_2(X_2) \tag{3}$$

where $\lambda$ is a parameter. It is worth pointing out that this assumption for a composed system with long-range interaction is reasonable in general. Take a self-gravitation system for example, as shown in Fig. 1, the total mass M is isotropically distributed in the system then the mass density can be easily written as

$$\rho = \frac{M}{\frac{4}{3}\pi R^3}. \tag{4}$$

And one can get the gravitational potential energy of such a system as

$$V_{total} = \int_0^R \frac{G(\rho \frac{4}{3}\pi r^3)dm}{r}, \tag{5}$$

where $G$ is the gravitational constant and

$$dm = \rho 4\pi r^2 dr. \tag{6}$$

Substituting Eq. (6) into Eq. (5) yields

$$V_{total} = G\rho^2 \frac{(4\pi)^2}{3}\int_0^R \frac{r^5 dr}{r} = G\rho^2 \frac{(4\pi)^2}{3}\frac{R^5}{5}. \tag{7}$$

If the above system was separated into two parts, as shown in Fig. 2. From Eq. (7) one can get the potential energy of part 1 directly,

$$V_1 = G\rho^2 \frac{(4\pi)^2}{3} \int_0^{R_1} \frac{r^5 dr}{r} = G\rho^2 \frac{(4\pi)^2}{3} \frac{R_1^5}{5}. \tag{8}$$

While the potential energy of part 2 can be written as

$$V_2 = G\rho^2 \frac{(4\pi)^2}{3} \int_{R_1}^{R} \frac{(r^3 - R_1^3) r^2 dr}{r} = G\rho^2 \frac{(4\pi)^2}{3} \left( \frac{R^5}{5} - \frac{R_1^3 R^2}{2} + \frac{3R_1^5}{10} \right). \tag{9}$$

From Eqs. (7)-(9) one can get

$$V_{total} = V_1 + V_2 + V_1 V_2 \frac{\frac{R_1^3 R^2}{2} - \frac{R_1^5}{2}}{\frac{R_1^5 R^5}{25} - \frac{R_1^8 R^2}{10} + \frac{3R_1^{10}}{50}}$$

$$= V_1 + V_2 + V_1 V_2 \frac{25}{R_1^5} \frac{(k^2 - 1)}{2k^5 - 5k^2 + 3}, \tag{10}$$

where $k = (R/R_1)$, the third item in the right hand side of Eq. (10) implies that the potential energy of a self-gravitating system is *nonadditive* since $\frac{(k^2 - 1)}{2k^5 - 5k^2 + 3} \neq 0$ unless $k \to \infty$. Comparing Eqs. (2) and (3) with Eq. (10) it's obvious that

$$\lambda = \frac{25}{R_1^5} \frac{(k^2 - 1)}{2k^5 - 5k^2 + 3} \neq 0.$$

3. **Probability distribution for a finite system with non-additive energy**

The following discussion is made along the line of the reference [6] without the hypothesis of thermodynamic limit ($N \to \infty$) and of additive energy. But it is still supposed that $H_1(X_1) \ll E$.

The probability distribution of $\Sigma_1$ is given by

$$p(X_1) = \frac{1}{\Omega(E)} \int_{(X_2)} \delta[E - H_1(X_1) - H_2(X_2) - \lambda H_1(X_1)H_2(X_2)]dX_2$$

$$= \frac{1}{\Omega(E)} \int_{(X_2)} \delta\{E - H_1(X_1) - [1 + \lambda H_1(X_1)][K(P_2) + V(R_2)]\}dP_2 dR_2$$

$$= \frac{1}{\Omega(E)} \int_{(R_2)} \Omega_k\{E - H_1(X_1) - [1 + \lambda H_1(X_1)]V(R_2)\}dR_2, \tag{11}$$

where $K(P_2)$ is the kinetic energy and $V(R_2)$ the potential energy of the particles in $\Sigma_2$, $P_2$ represents all their momenta and $R_2$ all their coordinates. $\Omega_k\{.\}$ is given by

$$\Omega_k\{y\} = \int_{(P_2)} \delta\{y - [1 + \lambda H_1(X_1)]K(P_2)\}dP_2 \tag{12}$$

$$= \int_{(P_2)} \delta[y - u(\lambda, H_1, P_2)]dP_2$$

with $y = E - H_1 - (1 + \lambda H_1)V(R_2)$ and $u(\lambda, H_1, P_2) = [1 + \lambda H_1(X_1)]K(P_2)$. $\Omega_k\{y\}$ is equal to the derivative of the volume of momentum space related to $P_2$ by the quantity $u(\lambda, H_1, P_2)$, enclosed within the hyper surface corresponding to $u(\lambda, H_1, P_2)$, i.e., $\Omega_k\{y\} = \partial\Gamma_k(y)/\partial y$ [6] where

$$\Gamma_k(y) = \int_{u \leq y} dP_2. \tag{13}$$

The quantity $u(\lambda, H_1, P_2)$, however, is equal to

$$u(\lambda, H_1, P_2) = [1 + \lambda H_1(X_1)] \sum_{n=1}^{N_2} \frac{\mathbf{P}_n^2}{2m_n}$$

$$= \sum_{n=1}^{N_2} \sum_{\alpha=1}^{3} \frac{(1 + \lambda H_1)}{2m_n}(\mathbf{P}_\alpha)_n^2. \tag{14}$$

Thus, if introducing the new variables,

$$D_k = \sqrt{\frac{1 + \lambda H_1}{2m_n}}(\mathbf{P}_\alpha)_n, \text{ where } k = 3(n-1) + \alpha, \tag{15}$$

we may write the equation for the hypersurface, corresponding to $u = y$ and enclosing the volume $\Gamma_k(y)$, in the form $\sum_{k=1}^{3N_2} D_k^2 = y$. According to some geometrical considerations [6], the above integration (13) gives

$$\Gamma_k(y) = a\, y^{3N_2/2}, \qquad (16)$$

where $a$ depends only on $N_2$. We finally obtain

$$\Omega_k\{y\} = b\, y^{3N_2/2-1}$$

$$= b[E - H_1 - (1+\lambda H_1)V]^{3N_2/2-1}, \qquad (17)$$

where $b = 3aN_2/2$. Then Eq.(11) becomes

$$p(X_1) = \frac{b}{\Omega(E)} \int_{(R_2)} \{E - H_1(X_1) - [1+\lambda H_1(X_1)]V(R_2)\}^{3N_2/2-1}\, dR_2. \qquad (18)$$

Up to now, we have not used any conditions of approximation for the above derivations so the equation (18) is exact. From Eq. (18) we can see that the form of potential of the thermostat, i.e. $V(R_2)$, the particles' number of the thermostat $N_2$ and the parameter $\lambda$ will affect the probability distribution function to some extent. On the other hand, the parameter $\lambda$ may depend on the energy of subsystems and some other physical quantities of the system. However, from Eqs. (5) to (10) we can see that $\lambda$ is the result of the integrals. $R_1$ and $R$ are lower and upper limits of the integrals and they are independent from the integral variable $R_2$ in Eq. (18), so $\lambda$ is also independent from $R_2$. Below we will discuss these cases in detail.

## 4. Discussions

By the mean field theory, all the interactions among the particles can be replaced by an average or effective interaction. This is a mathematical simplification for a system with complex interactions including the long-range one. It's reasonable to consider the potential energy of each particle in $\sum_2$ as a constant, i.e., $V(R_2) = C$. Substitute it into Eq. (18) one can get

$$p(X_1) = \frac{b}{\Omega(E)} \int_{(R_2)} \{E - H_1(X_1) - C[1 + \lambda H_1]\}^{3N_2/2-1} dR_2$$

$$= \frac{b}{\Omega(E)} (E-C)^{\frac{3N_2}{2}-1} \left[1 - \frac{(1+C\lambda)H_1}{E-C}\right]^{\frac{3N_2}{2}-1} \int_{(R_2)} dR_2$$

$$= \frac{b\Omega_2}{\Omega(E)} (E-C)^{\frac{3N_2}{2}-1} \left[1 - \frac{(1+C\lambda)H_1}{E-C}\right]^{\frac{3N_2}{2}-1}. \tag{19}$$

Eq. (19) can be written as

$$p(X_1) = C_2 \left[1 - \frac{(1+C\lambda)H_1}{E-C}\right]^{\frac{3N_2}{2}-1}, \tag{20}$$

where $C_2$ is a normalization constant. On the other hand we have already agreed to assume that $N_1 \ll N_2$, we may also get

$$E - C \approx \frac{3}{2} N_2 \Theta', \tag{21}$$

where $\frac{1}{2}\Theta'$ is the mean kinetic energy per degree of freedom of $\sum_2$, it has directly association with the physical temperature of the system. Substituting Eq. (21) into (20) we can get

$$p(X_1) = C_2 \left[1 - \frac{(1+C\lambda)H_1}{\frac{3}{2}N_2\Theta'}\right]^{\frac{3N_2}{2}-1} \approx C_2 \left[1 - \frac{(1+C\lambda)H_1}{\left(\frac{3}{2}N_2-1\right)\Theta'}\right]^{\frac{3N_2}{2}-1}$$

$$= C_2 \left(1 - (1-q)\frac{(1+C\lambda)H_1}{\Theta'}\right)^{\frac{1}{(1-q)}}. \tag{22}$$

Under the limit of $C \to 0$ Eq. (22) can be written as

$$p(X_1) = C_2 \left(1 - (1-q)\frac{H_1(X_1)}{\Theta'}\right)^{\frac{1}{(1-q)}}, \tag{23}$$

where $q = \frac{3N_2 - 4}{3N_2 - 2}$ [9] and $C_1$ is a normalization constant. This is the conclusion of Ref. [6]. From Eq. (23) it's obvious that for an ideal finite system the probability distribution is in q-exponential form. When the particle number of $\sum_2$ tends to infinite, i.e., $q \to 1$, the distribution function will reduce to exponential one under thermodynamic limit.

It is seen from Eq. (22) that the probability distribution of the finite system with nonadditive energy is dependent not only on the number of particles of the thermostat $N_2$ (or $q$) but also on the parameter $C\lambda$ which describes the energy's nonextensivity of the system. For any given $N_2$ one can generate the curves of $p_1/C_2$ varying with $H_1/\Theta'$, as shown in Fig. 3. It is worth to note that some interesting results can be deduced from the curves in Fig.3. (i) The probability distribuiton for the finite system with nonadditive energy decreases with the increase of the parameters $C\lambda$ at all $H_1/\Theta'$. (ii) The probability distribution is a monotonically decreasing function of Hamiltonian, which is similar to the case of ideal gas. (iii) The $p_1|_{H_1=0}$ and $p_1|_{H_1 \to \infty}$ will not change with different values of parameter $C\lambda$, so the difference of probability distribution with different $C\lambda$ first increases then decreases with the increasing of $H_1/\Theta'$ and there exists a maximal difference.

When $N_2 \to \infty$, i.e. $q = \dfrac{3N_2 - 4}{3N_2 - 2} \to 1$, all the conclusions above will reduce to the thermodynamic limit. The curve of $C\lambda = 0$ is in accordance with the exponential function. This kind of distribution has translation invariance [14], which means the distribution function will keep invariant if the Hamiltonian of the system takes a spectrum shift. This shift can be caused by a constant external potential which has an arbitrary value (zero or nonzero does not make any difference). However, the interaction potential in the system is different from the external one. From Eq. (22) we can find $p(X_1) \propto Exp[-(1+C\lambda)H_1/\Theta']$ at the $q \to 1$ limit, which has neither translation invariance nor scale invariance [14]. So the difference between additive energy ($C\lambda = 0$) and nonadditive energy ($C\lambda \neq 0$) can not be neglected. This point can be clearly seen from Fig. 4. The nonextensivity of the internal energy will distort the probability distribution from the exponential one ($C\lambda = 0$) even for a large heat bath.

In fact the potential energy of $\sum_2$, i.e., $V(R_2)$ is a function of $R_2$. The concrete form of $V(R_2)$ depends on the interactions between particles in $\sum_2$. However substituting $V(R_2)$ as function of $R_2$ into the calculation will cause some mathematical difficulties; it's then still an open question.

## 5. Conclusions

In summary, the canonical distribution for finite systems in equilibrium is studied in the present work. Due to the long-range interaction between the system and the thermostat a very simple model is presented to illustrate the nonadditive energy of the system. Based on Eq. (3) we analyze the possible canonical distributions for different interactions. The nonextensivity of the energy of the system greatly influences the distribution function whether the system is finite or not. The deviations of the probability distribution from the ideal case always increase with the increasing energy nonextensivity (parameter $C\lambda$) of the system. When the interactions among the system tend to zero ($C\lambda = 0$) the distribution function will reduce to a q-exponential function, i.e. Eq. (23). It's also shown that the distribution function for a finite system within long-range interaction can be presented in a q-exponential form, where the parameter q has a directly correlation with the particle number of the thermostat, i.e. $q = \frac{3N_2 - 4}{3N_2 - 2}$. It's naturally that the distribution function will reduce to the exponential form at the thermodynamic limit ($N_2 \to \infty$). The results of the present work is general, it's expounded that the q-exponential distribution can be used to describe the finite system in thermodynamic equilibrium, also the results of ideal finite system and the thermodynamic limit can be considered as special cases of our framework.

Acknowledgements

This work has been supported by the region des Pays de la Loire of France under grant number 2007-6088 and by the Science Research Fund (No. 07BS105), Huaqiao University, People's Republic of China.

Fig captions:

Fig. 1: A spherical self-gravitation system with radius R and mass M (isotropy distributed). The gravitational potential energy of the total system can be calculated by integral method, and the integral can be considered between a spherical part with radius *r* and the spherical shell with thickness *dr*.

Fig. 2: The spherical self-gravitation system is separated into two subsystems by a spherical surface with radius $R_1$. Each part has a gravitational potential energy respectively. The calculations of these two potential energies are the same as the previous one.

Fig. 3: The curves of the probability distribution varying with the Hamiltonian for some different values of Cλ at $N_2 = 100$, $C\lambda$ is the nonextensivity measurement of the energy of finite system. $C\lambda = 0$ represents the ideal case which is described in Ref. [6]. $C\lambda > 0$ means the long-range interaction among subsystems is attractive while $C\lambda < 0$ means exclusive.

Fig. 4: The curves of the probability distribution varying with the Hamiltonian for some different values of Cλ at $N_2 = 10^5$. The meaning of parameter Cλ is the same as the one in Fig. 3.

Fig.1

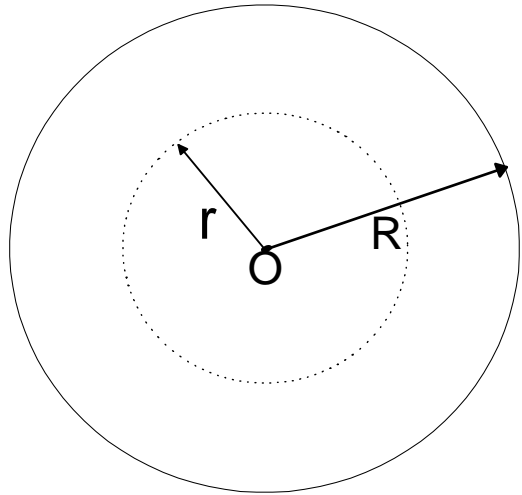

Fig.2

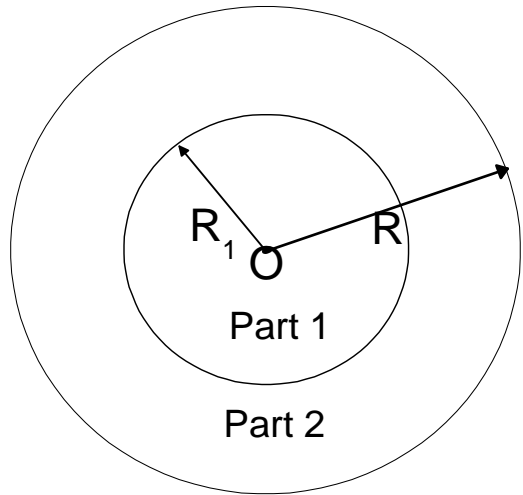

Fig. 3

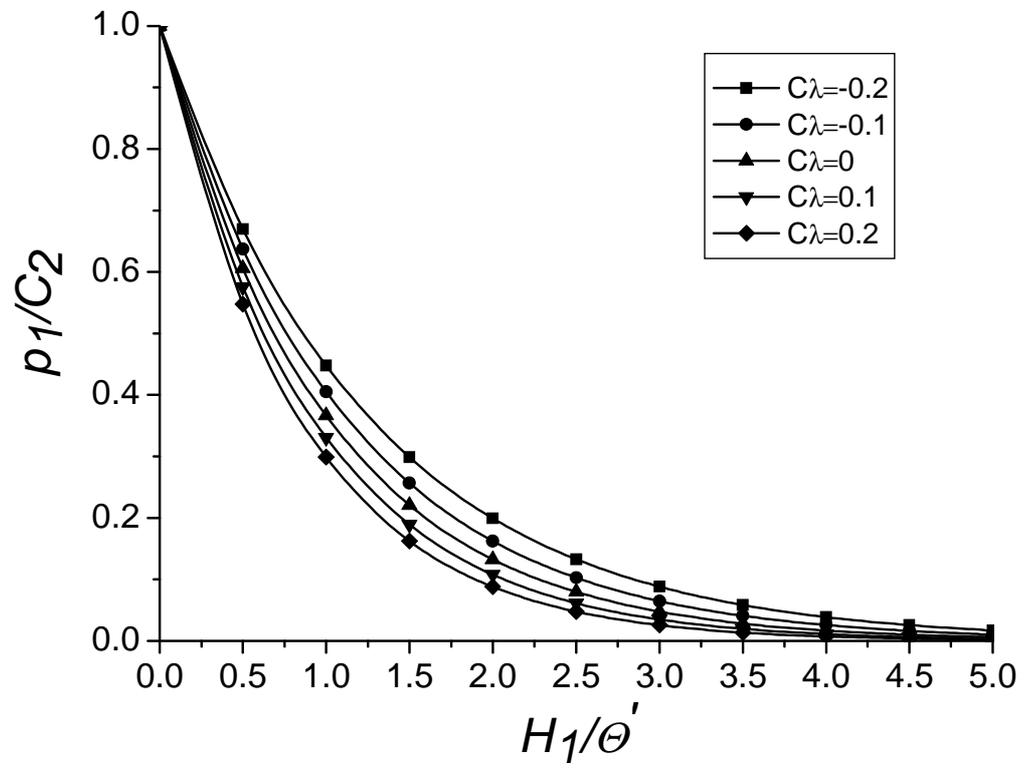

Fig. 4

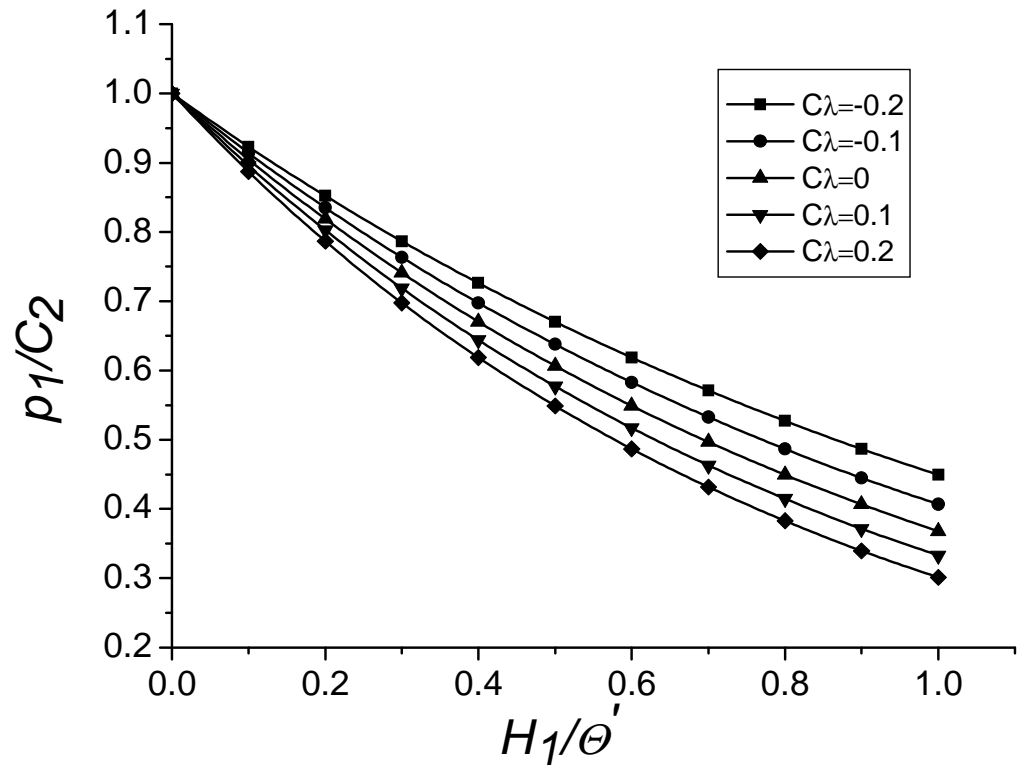